\documentclass[prd,onecolumn,nofootinbib,showpacs,showkeys]{revtex4}
\newcommand\gothfamily{\usefont{U}{ygoth}{m}{n}}
\DeclareTextFontCommand{\textgoth}{\gothfamily}

\begin{document}

\title{TORSION AS ELECTROMAGNETISM AND SPIN}

\author{{\bf Nikodem J. Pop\l awski}}

\affiliation{Department of Physics, Indiana University, Swain Hall West, 727 East Third Street, Bloomington, IN 47405, USA}
\email{nipoplaw@indiana.edu}

\noindent
{\em International Journal of Theoretical Physics}\\
Vol. {\bf 49}, No. 7 (2010) 1481--1488\\
\copyright\,Springer Science+Business Media, LLC
\vspace{0.4in}

\begin{abstract}
We show that it is possible to formulate the classical Einstein-Maxwell-Dirac theory of spinors interacting with the gravitational and electromagnetic fields as the Einstein-Cartan-Kibble-Sciama theory with the Ricci scalar of the traceless torsion, describing gravity, and the torsion trace acting as the electromagnetic potential.
\end{abstract}


\maketitle

\section{Sourceless Maxwell Equations}
Consider the Einstein-Cartan Lagrangian density with additional terms containing torsion \cite{Hehl,torsion1,torsion2}:
\begin{equation}
{\cal L}=-\frac{1}{2\kappa}\sqrt{-g}R_{ik}g^{ik}-\frac{\alpha}{4}\sqrt{-g}Q_{ik}Q^{ik}+\frac{\beta}{2\kappa}\sqrt{-g}S_i S^i,
\label{L}
\end{equation}
where $R_{ik}$ is the Ricci tensor of the affine connection $\Gamma^{\,\,k}_{i\,j}$,
\begin{equation}
S_i=S^k_{\phantom{k}ik}
\end{equation}
is the torsion vector (trace),
\begin{equation}
Q_{ik}=S_{k,i}-S_{i,k}
\label{tt}
\end{equation}
has the form of the electromagnetic field tensor with $S_i$ representing the electromagnetic potential \cite{Bor} (comma denotes a partial derivative), and $\alpha>0$.
Substituting
\begin{equation}
\Gamma^{\,\,k}_{i\,j}=\{^{\,\,k}_{i\,j}\}+C^k_{\phantom{k}ij},
\label{conn}
\end{equation}
where $\{^{\,\,k}_{i\,j}\}$ are the Christoffel symbols and
\begin{equation}
C^k_{\phantom{k}ij}=S^k_{\phantom{k}ij}+S_{ij}^{\phantom{ij}k}+S_{ji}^{\phantom{ji}k}=-C^{\phantom{i}k}_{i\phantom{k}j}
\label{cont}
\end{equation}
is the contortion tensor, into the definition of the Ricci scalar $R$ gives \cite{Scho}
\begin{equation}
R=R^{\{\}}+C^{ik}_{\phantom{ik}k:i}-C^{ik}_{\phantom{ik}i:k}+C^{jk}_{\phantom{jk}k}C^i_{\phantom{i}ji}-C^{jk}_{\phantom{jk}i}C^i_{\phantom{i}jk},
\label{Ricci}
\end{equation}
where $R^{\{\}}$ is the Riemannian Ricci scalar and : denotes a covariant derivative with respect to $\{^{\,\,k}_{i\,j}\}$.

The second and third term on the right-hand side of (\ref{Ricci}) multiplied by $\sqrt{-g}$ are total covariant divergences, so they do not contribute to the field equations.
The variation of $\sqrt{-g}R$ with respect to the contortion (which is equivalent to varying with respect to the torsion), that contributes to the field equations, is then
\begin{equation}
\delta(\sqrt{-g}R)=\sqrt{-g}(C^m_{\phantom{m}lm}g^{ik}+C^i_{\phantom{i}mn}g^{mn}\delta^k_l-C^k_{\phantom{k}lm}g^{im}-C^i_{\phantom{i}ml}g^{mk})\delta C^l_{\phantom{l}ik}.
\end{equation}
The variation of $\sqrt{-g}Q_{ik}Q^{ik}$ with respect to $C$ is
\begin{equation}
\delta(\sqrt{-g}Q_{ik}Q^{ik})=4\sqrt{-g}Q^{ik}\delta S_{k,i}=-4(\sqrt{-g}Q^{ik})_{,i}\delta S_k=-2(\sqrt{-g}Q^{ji})_{,j}\delta^k_l\delta C^l_{\phantom{l}ik}=-2\sqrt{-g}Q^{ji}_{\phantom{ji}:j}\delta^k_l\delta C^l_{\phantom{l}ik},
\end{equation}
where we omit a total divergence.
The variation of $\sqrt{-g}S_i S^i$ with respect to $C$ is
\begin{equation}
\delta(\sqrt{-g}S_i S^i)=\sqrt{-g}S^i\delta^k_l\delta C^l_{\phantom{l}ik}.
\end{equation}
Therefore the variation of ${\cal L}$ with respect to $C^l_{\phantom{l}ik}$ is
\begin{equation}
\delta{\cal L}=\sqrt{-g}\biggl(-\frac{1}{2\kappa}(C^m_{\phantom{m}lm}g^{ik}+C^i_{\phantom{i}mn}g^{mn}\delta^k_l-C^k_{\phantom{k}lm}g^{im}-C^i_{\phantom{i}ml}g^{mk})+\frac{\alpha}{2}Q^{ji}_{\phantom{ji}:j}\delta^k_l+\frac{\beta}{2\kappa}S^i\delta^k_l\biggr)\delta C^l_{\phantom{l}ik},
\end{equation}
and the resulting field equation $\delta{\cal L}/\delta C^l_{\phantom{l}ik}=0$ gives
\begin{equation}
\delta^k_{[i} C^m_{\phantom{m}l]m}+\delta^k_{[l} C_{i]m}^{\phantom{i]m}m}-C^k_{\phantom{k}[li]}+C^k_{\phantom{k}[il]}-\alpha\kappa\delta^k_{[l} Q^j_{\phantom{j}i]:j}-\beta\delta^k_{[l}S_{i]}=0,
\end{equation}
where $[\,]$ denotes antisymmetrization.
Contracting the indices $k,l$ gives
\begin{equation}
\alpha\kappa Q^j_{\phantom{j}i:j}+\biggl(\beta+\frac{8}{3}\biggr)S_i=0,
\label{mas}
\end{equation}
so the vector field $S_i$ in the Lagrangian density (\ref{L}) is massless and satisfies Maxwell-like equations if \cite{Ham}
\begin{equation}
\beta=-\frac{8}{3}.
\end{equation}
Accordingly, the Lagrangian density (\ref{L}) is
\begin{equation}
{\cal L}=-\frac{1}{2\kappa}\sqrt{-g}\biggl(R+\frac{8}{3}S_i S^i\biggr)-\frac{\alpha}{4}\sqrt{-g}Q_{ik}Q^{ik}.
\end{equation}

\section{Maxwell Equations with Spinor Sources}
The curvature tensor with two Lorentz and two coordinate indices depends on the spin connection $\omega^a_{\phantom{a}bi}=-\omega_{b\phantom{a}i}^{\phantom{b}a}$ and its first derivatives \cite{Hehl}:
\begin{equation}
R^a_{\phantom{a}bij}=\omega^a_{\phantom{a}bj,i}-\omega^a_{\phantom{a}bi,j}+\omega^a_{\phantom{a}ci}\omega^c_{\phantom{c}bj}-\omega^a_{\phantom{a}cj}\omega^c_{\phantom{c}bi}.
\label{curva}
\end{equation}
The double contraction of the tensor (\ref{curva}) with the tetrad $e^i_a$ gives the Ricci scalar
\begin{equation}
R=R^a_{\phantom{a}bij}e^i_a e^{bj}.
\end{equation}
The torsion vector is related to the spin connection by
\begin{equation}
S_i=\omega^k_{\phantom{k}[ik]}+e^a_{[i,k]}e^k_a,
\end{equation}
and the contortion is
\begin{equation}
C_{ijk}=\omega_{ijk}+e_{ia}e^a_{[j,k]}-e_{ja}e^a_{[i,k]}-e_{ka}e^a_{[i,j]}.
\label{contor}
\end{equation}

Consider the Lagrangian density
\begin{equation}
{\cal L}=-\frac{1}{2\kappa}e\biggl(R+\frac{8}{3}S_i S^i\biggr)-\frac{\alpha}{4}eQ_{ij}Q^{ij}+{\cal L}_m,
\label{Ls}
\end{equation}
where $e=\mbox{det}(e_i^a)=\sqrt{-g}$ and ${\cal L}_m$ is a Lagrangian density for matter.
The variation of ${\cal L}_m$ with respect to the spin connection defines the spin density $\Sigma_{ab}^{\phantom{ab}i}$:
\begin{equation}
\delta{\cal L}_m=\frac{1}{2}\Sigma_{ab}^{\phantom{ab}i}\delta \omega^{ab}_{\phantom{ab}i}=\frac{1}{2}\Sigma_{jk}^{\phantom{jk}i}\delta C^{jk}_{\phantom{jk}i}.
\end{equation}
The gravitational part of ${\cal L}$ (the Einstein-Cartan-Kibble-Sciama Lagrangian density \cite{sptor1,sptor2}) can be written as
\begin{equation}
eR=2E^{ik}_{ab}(\omega^{ab}_{\phantom{ab}k,i}+\omega^a_{\phantom{a}ci}\omega^{cb}_{\phantom{cb}k}),
\end{equation}
where $E^{ik}_{ab}=ee^{[i}_a e^{k]}_b$.

The field equation $\delta{\cal L}/\delta\omega^l_{\phantom{l}ik}=0$ gives a relation between the spin connection and spin density \cite{Hehl}.
The variation of $eR$ with respect to the spin connection is
\begin{equation}
\delta(eR)=2E^{ik}_{ab}\delta(\omega^{ab}_{\phantom{ab}k,i}+\omega^a_{\phantom{a}ci}\omega^{cb}_{\phantom{cb}k})=2(E^{ik}_{ab,k}+E^{ik}_{ac}\omega_{b\phantom{c}k}^{\phantom{b}c}-E^{ik}_{cb}\omega^c_{\phantom{c}ak})\delta\omega^{ab}_{\phantom{ab}i}.
\label{var}
\end{equation}
A total (with respect to both the affine and spin connection) covariant derivative $|$ of the tensor density $E^{ik}_{ab}$ vanishes due to vanishing of the total covariant derivative of the tetrad:
\begin{equation}
E^{ik}_{ab|k}=E^{ik}_{ab,k}-\omega^c_{\phantom{c}ak}E^{ik}_{cb}-\omega^c_{\phantom{c}bk}E^{ik}_{ac}+S^i_{\phantom{i}jk}E^{jk}_{ab}+\Gamma^{\,\,j}_{k\,j}E^{ik}_{ab}-\Gamma^{\,\,j}_{j\,k}E^{ik}_{ab}=0,
\end{equation}
so (\ref{var}) gives, omitting total divergences,
\begin{equation}
\delta(eR)=-2(S^i_{\phantom{i}lk}E^{lk}_{ab}+2S_kE^{ik}_{ab})\delta\omega^{ab}_{\phantom{ab}i}=-2e(S^{kij}-S^i g^{jk}+S^j g^{ik})\delta\omega_{ijk}.
\end{equation}
The variation of $eQ_{ij}Q^{ij}$ with respect to $\omega$ is
\begin{equation}
\delta(eQ_{ij}Q^{ij})=4eQ^{ij}\delta\omega^k_{\phantom{k}[jk],i}\rightarrow-2(eQ^{ij})_{,i}\delta\omega^k_{\phantom{k}jk}=-2e(Q^{l[j})_{:l}g^{i]k}\delta\omega_{ijk}.
\end{equation}
The variation of $eS_i S^i$ with respect to $\omega$ is
\begin{equation}
\delta(eS_i S^i)=eS^j\delta\omega^k_{\phantom{k}jk}=eS^{[j} g^{i]k}\delta\omega_{ijk}.
\end{equation}
Consequently, $\delta{\cal L}/\delta\omega^l_{\phantom{l}ik}=0$ gives the relation between the torsion tensor and the spin density:
\begin{equation}
S^{kij}-\frac{1}{3}(S^i g^{jk}-S^j g^{ik})-\frac{\alpha\kappa}{4}(Q^{li}_{\phantom{li}:l}g^{jk}-Q^{lj}_{\phantom{lj}:l}g^{ik})+\frac{\kappa}{2e}\Sigma^{ijk}=0.
\label{field}
\end{equation}

Contracting the field equation (\ref{field}) with respect to the indices $j,k$ brings it into the Maxwell-like form:
\begin{equation}
\frac{3\alpha}{4}Q^{li}_{\phantom{li}:l}=\frac{1}{2e}\Sigma^{ik}_{\phantom{ik}k}.
\label{Max}
\end{equation}
Substituting (\ref{Max}) into (\ref{field}) gives
\begin{equation}
S^{kij}-\frac{1}{3}(S^i g^{jk}-S^j g^{ik})+\frac{\kappa}{2e}\biggl(\Sigma^{ijk}+\frac{2}{3}g^{k[i}\Sigma^{j]l}_{\phantom{j]l}l}\biggr)=0,
\label{EC}
\end{equation}
from which it follows that the totally antisymmetrized torsion tensor is proportional to the totally antisymmetrized spin density:
\begin{equation}
S^{[ijk]}=-\frac{\kappa}{2e}\Sigma^{[ijk]}.
\label{alg}
\end{equation}
Taking the trace of both sides of (\ref{EC}) gives the identity.
However, the trace of the torsion is already related to the trace of the spin density via (\ref{tt}) and (\ref{Max}), leading to a second-order differential equation for the torsion trace.
The Einstein equations relating the Ricci tensor to the energy-momentum tensor result from varying the Lagrangian density (\ref{Ls}) with respect to the tetrad $e^i_a$.
The contracted Bianchi identities applied to the Einstein equations are consistent with (\ref{field}).

The Dirac Lagrangian density for a spinor field $\psi$ with mass $m$ in the presence of the gravitational field is
\begin{equation}
\textgoth{L}_m=\frac{ie}{2}(\bar{\psi}\gamma^i \psi_{|i}-\bar{\psi}_{|i}\gamma^i \psi)-em\bar{\psi}\psi,
\label{Dir1}
\end{equation}
where $\bar{\psi}$ is the adjoint spinor corresponding to $\psi$ and $\hbar=c=1$.
Substituting the definition of a covariant derivative of a spinor,
\begin{equation}
\psi_{|i}=\psi_{,i}-\Gamma_i\psi,\,\,\,\bar{\psi}_{|i}=\bar{\psi}_{,i}+\bar{\psi}\Gamma_i,
\label{covdev}
\end{equation}
into (\ref{Dir1}) gives
\begin{equation}
\textgoth{L}_m=\frac{ie}{2}(\bar{\psi}\gamma^i \psi_{,i}-\bar{\psi}_{,i}\gamma^i \psi)-\frac{ie}{2}\bar{\psi}\{\gamma^i,\Gamma_i\}\psi-em\bar{\psi}\psi,
\label{Dir2}
\end{equation}
where $\{,\}$ denotes anticommutation.
The spinor connection $\Gamma_i$ is given by the condition $\gamma^a_{\phantom{a}|i}=\omega^a_{\phantom{a}bi}\gamma^b-[\Gamma_i,\gamma^a]=0$ (where $[,]$ denotes commutation) \cite{Hehl}:
\begin{equation}
\Gamma_i=-\frac{1}{4}\omega_{abi}\gamma^a \gamma^b+V_i.
\label{FI}
\end{equation}
The first term on the right-hand side of (\ref{FI}) is referred to as the Fock-Ivanenko coefficients $\Gamma_i^{(FI)}$ \cite{sptor1,sptor2} and $V_i$ is an arbitrary vector multiple of the unit matrix.
The relation $\bar{\psi}=\psi^\dagger\gamma^0$ implies that $V_i$ is imaginary.

If $V_k$ is to be a purely geometric quantity, it must be proportional to a vector constructed from the torsion and/or curvature.
The simplest possibility for such a vector is the torsion trace:
\begin{equation}
V_k=-iq\sqrt{\alpha}S_k,
\end{equation}
where $q$ is a real scalar.
Therefore (\ref{Dir2}) becomes
\begin{equation}
\textgoth{L}_m=\frac{ie}{2}(\bar{\psi}\gamma^i \psi_{,i}-\bar{\psi}_{,i}\gamma^i \psi)+\frac{ie}{8}\omega_{abi}\bar{\psi}\{\gamma^i,\gamma^a \gamma^b\}\psi-e\sqrt{\alpha}S_i j^i-em\bar{\psi}\psi,
\label{Dir3}
\end{equation}
where
\begin{equation}
j^i=q\bar{\psi}\gamma^i\psi.
\end{equation}
The spin density corresponding to the Lagrangian density (\ref{Dir3}) is
\begin{equation}
\Sigma^{ijk}=\frac{ie}{2}\bar{\psi}\gamma^{[i}\gamma^j\gamma^{k]}\psi+e\sqrt{\alpha}j^{[i}g^{j]k},
\label{spin}
\end{equation}
so its trace is
\begin{equation}
\Sigma^{ik}_{\phantom{ik}k}=\frac{3}{2}\sqrt{\alpha}ej^i.
\label{trace}
\end{equation}
Substituting (\ref{trace}) into (\ref{Max}) gives
\begin{equation}
\sqrt{\alpha}Q^{ki}_{\phantom{ki}:k}=j^i.
\label{ME}
\end{equation}

If we, following Borchsenius \cite{Bor}, associate $\sqrt{\alpha}S_i$ with the electromagnetic potential $A_i$ then $\sqrt{\alpha}Q_{ij}$ represents the electromagnetic field tensor $F_{ij}$ and the term $-\frac{\alpha}{4}eQ_{ij}Q^{ij}$ in (\ref{Ls}) corresponds to the Maxwell Lagrangian density $-\frac{1}{4}eF_{ij}F^{ij}$.
Accordingly, $V_k=-iqA_k$ coincides with the electromagnetic $U(1)$-covariant derivative in (\ref{covdev}) and (\ref{FI}), $q$ is the electric charge of a spinor $\psi$ (in units of the charge of the electron) and $j^i$ is the covariantly conserved ($j^i_{\phantom{i}:i}=0$) electromagnetic current related to this spinor.
The field equations (\ref{ME}) reproduce the Maxwell equations.
Equation (\ref{EC}) becomes
\begin{equation}
S^{kij}=\frac{1}{3\sqrt{\alpha}}(A^i g^{jk}-A^j g^{ik})-\frac{\kappa}{2e}\biggl(\Sigma^{ijk}+\frac{2}{3}g^{k[i}\Sigma^{j]l}_{\phantom{j]l}l}\biggr).
\label{ECD}
\end{equation}
Substituting (\ref{spin}) into (\ref{ECD}) gives
\begin{equation}
S^{kij}=\frac{1}{3\sqrt{\alpha}}(A^i g^{jk}-A^j g^{ik})-\frac{i\kappa}{4}\bar{\psi}\gamma^{[i}\gamma^j\gamma^{k]}\psi.
\label{tors}
\end{equation}

The torsion tensor can be decomposed into three irreducible parts: the trace, axial trace (which is dual to the totally antisymmetric component) and trace-free part \cite{Hehl}.
It follows from (\ref{tors}) that the torsion trace represents the electromagnetic potential, the torsion axial trace corresponds to the Dirac spin pseudovector, and the trace-free part of the torsion vanishes.
Similarly, the spin density can be decomposed into the same three irreducible parts because it has the same symmetry properties (except the irrelevant order of the indices).
It follows from (\ref{spin}) that the trace of the spin density represents the Dirac electromagnetic current, the axial trace of the spin density corresponds to the Dirac spin pseudovector, and the trace-free part of the spin density vanishes.
The trace parts of the torsion tensor and spin density are related to one another differentially via the Maxwell equations (\ref{Max}) (in the original Einstein-Cartan theory with Dirac sources the two traces vanish).
The axial trace parts of the torsion tensor and spin density are related to one another algebraically via (\ref{alg}), as in the Einstein-Cartan theory.
This construction does not lead to contradictions because the trace and axial trace are irreducible components of the respective tensor quantities.
Thus the relation between the trace parts of $S_{ijk}$ and $\Sigma_{ijk}$ is independent of the relation between the axial trace parts of $S_{ijk}$ and $\Sigma_{ijk}$.

\section{Dirac Equation}
The field equations $\delta{\cal L}/\delta\psi=0$ and $\delta{\cal L}/\delta\bar{\psi}=0$ give the (equivalent) equations for spinor fields.
The variation of the Dirac Lagrangian density (\ref{Dir1}) with respect to $\bar{\psi}$ gives, after omitting total divergences,
\begin{equation}
\frac{i}{2}\bigl(e\gamma^k\psi_{,k}+(e\gamma^k\psi)_{,k}-e\{\Gamma_k^{(FI)},\gamma^k\}\psi\bigr)-eq\sqrt{\alpha}S_i\gamma^i\psi-em\psi=0.
\label{HI1}
\end{equation}
Substituting
\begin{equation}
(e\gamma^k\psi)_{,k}=e\gamma^k\psi_{,k}+e\gamma^k_{\phantom{k};k}\psi-2eS_k\gamma^k\psi=e\gamma^k\psi_{,k}+e[\Gamma_k,\gamma^k]\psi-2eS_i\gamma^i\psi,
\end{equation}
where ; denotes a covariant derivative with respect to $\Gamma^{\,\,k}_{i\,j}$, into (\ref{HI1}) gives
\begin{equation}
i\gamma^k\psi_{,k}-i\gamma^k\Gamma_k^{(FI)}\psi-iS_k\gamma^k\psi-q\sqrt{\alpha}S_k\gamma^k\psi-m\psi=i\gamma^k(\psi_{|k}+iq\sqrt{\alpha}S_k\psi)-iS_k\gamma^k\psi-m\psi=0.
\label{HI2}
\end{equation}
Equations (\ref{conn}) and (\ref{contor}) give
\begin{equation}
C_{ijk}=\omega_{ijk}-\omega_{ijk}^{\{\}},
\end{equation}
where $\omega_{ijk}^{\{\}}$ is the spin connection corresponding to the Levi-Civita connection $\{^{\,\,k}_{i\,j}\}$.
Therefore
\begin{equation}
\psi_{|k}=\psi_{|k}^{\{\}}+\frac{1}{4}C_{ijk}\gamma^i\gamma^j\psi,
\end{equation}
where $\psi_{|k}^{\{\}}$ is a covariant derivative of a spinor with respect to the Levi-Civita connection.
Equations (\ref{cont}) and (\ref{tors}) give
\begin{equation}
\frac{1}{4}C_{ijk}\gamma^k\gamma^i\gamma^j-S_k\gamma^k=-\frac{i\kappa}{16}\bar{\psi}\gamma_{[i}\gamma_j\gamma_{k]}\psi\gamma^i\gamma^j\gamma^k,
\end{equation}
so (\ref{HI2}) becomes the Heisenberg-Ivanenko equation with the electromagnetic coupling \cite{sptor1,sptor2}:
\begin{equation}
i\gamma^k(\psi_{|k}^{\{\}}+iq\sqrt{\alpha}S_k\psi)+\frac{3\kappa}{8}(\bar{\psi}\gamma_k\gamma^5\psi)\gamma^k\gamma^5\psi=m\psi.
\label{HI3}
\end{equation}
The first term on the left-hand side of (\ref{HI3}) corresponds to the general-relativistic interaction of $\psi$ with the electromagnetic potential.
The second term, nonlinear in spinors, describes the Heisenberg-Ivanenko spinor self-interaction that introduces deviations from the Dirac equation at energies on the order of the Planck energy \cite{sptor1,sptor2}.

\section{Gauge Invariance}
Consider a transformation
\begin{equation}
S_{kij}\rightarrow S_{kij}+\frac{1}{3}(\lambda_{,i}g_{jk}-\lambda_{,j}g_{ik}),
\label{tran}
\end{equation}
where $\lambda$ is a scalar.
Although this transformation looks like the antisymmetric part of Einstein's $\lambda$-transformation of the affine connection, $\Gamma^{\,\,k}_{i\,j}\rightarrow \Gamma^{\,\,k}_{i\,j}-\frac{2}{3}\lambda_{,j}\delta^k_i$, we only transform the torsion tensor.
Consequently, the torsion vector transforms like the electromagnetic potential under a $U(1)$ gauge transformation \cite{Bor}:
\begin{equation}
S_i\rightarrow S_i+\lambda_{,i},
\end{equation}
and the contortion tensor changes as
\begin{equation}
C_{kij}\rightarrow C_{kij}+\frac{2}{3}(\lambda_{,i}g_{jk}-\lambda_{,k}g_{ji}).
\end{equation}
Therefore the transformation (\ref{tran}) is a geometric representation of the gauge transformation in electromagnetism.
The field equation (\ref{field}) is invariant under (\ref{tran}) because the traceless part of the torsion tensor,
\begin{equation}
S_{kij}^{(t)}=S_{kij}-\frac{1}{3}(S_i g_{jk}-S_j g_{ik}),
\end{equation}
is invariant under (\ref{tran}).
Accordingly, the Lagrangian density (\ref{Ls}) is also invariant under (\ref{tran}), which results from the invariance of $Q_{ij}$, because $R+\frac{8}{3}S_i S^i$ is equal to the Ricci scalar in which the torsion tensor is replaced by its traceless part:
\begin{equation}
R(S_{ijk})+\frac{8}{3}S_i S^i=R(S_{ijk}^{(t)}),
\end{equation}
and because the term $eS_i j^i$ in (\ref{Dir3}) changes under (\ref{tran}) by a total divergence since the current $j^i$ is conserved.
The gravitational field is thus described in the Lagrangian density by the Ricci scalar of the traceless torsion and the electromagnetic field is described by the square of the curl of the torsion trace.
If $\beta\neq-\frac{8}{3}$ in the Lagrangian density (\ref{L}) then the field equation (\ref{mas}) describes a massive vector field without gauge invariance because the gravitational part of (\ref{L}) depends on the full torsion tensor and not on its traceless part only.

\section{Concluding Remarks}
This paper shows that it is possible to formulate the classical Einstein-Maxwell-Dirac theory of spinors interacting with the gravitational and electromagnetic fields as the Einstein-Cartan-Kibble-Sciama theory \cite{Hehl} with the Ricci scalar of the traceless torsion, describing gravity, and the torsion trace acting as the electromagnetic potential (so the curl of the torsion trace represents the electromagnetic field tensor) \cite{Bor}.
One can construct a geometric formulation of the Einstein-Maxwell theory, where the segmental (homothetic) curvature tensor \cite{Scho} represents the electromagnetic field tensor and the trace of the nonmetricity tensor (the Weyl vector) acts like the electromagnetic potential, or even where the electromagnetic potential corresponds to a linear combination of the torsion and nonmetricity traces \cite{geom1,geom2,geom3,geom4,PO1,PO2}.
We favor the construction with the torsion vector representing the electromagnetic potential because torsion, unlike nonmetricity, has a geometrical meaning \cite{torsion1,torsion2}.
We extended the Einstein-Maxwell theory with the torsion vector representing the electromagnetic potential \cite{Bor,Ham} to spinors, as it has been done for the Einstein-Maxwell theory with the Weyl vector acting like the electromagnetic potential \cite{PO1,PO2}.

The geometrical theory of electromagnetism presented in this paper reproduces the Einstein-Maxwell theory in an elegant fashion, but it does not introduce new physics that is different from the Einstein-Cartan-Kibble-Sciama theory.
For example, spinors still obey the Heisenberg-Ivanenko equation which differs significantly from the general-relativistic Dirac eqution only at energies on the order of the Planck energy.
Relating electromagnetism to torsion, which is the tensorial part of the affine connection, seems natural, however.
In the presence of the gravitational field we generalize a partial derivative into a coordinate-covariant derivative by introducing the affine connection, while in the presence of the electromagnetic field we generalize it into a $U(1)$-covariant derivative by introducing the electromagnetic potential.
This relation may suggest the correct way of quantizing the gravitational field, since we already have a highly successful quantum theory of the electromagnetic field (QED).
We also note that massive vectors, characteristic to weak interactions, can be generated in the Einstein-Cartan gravity with additional terms containing torsion.
A successful theory unifying the gravitational and electromagnetic interactions on the classical level should be regarded as the classical limit of the quantum theory of all interactions, giving insights on how to construct such a theory, possibly with geometrization of spinor fields.
Therefore classical unified field theory is still a topic worthy of investigation.

\end{document}